\documentclass[10pt, doublecolumns]{IEEEtran}

\ifCLASSINFOpdf

\else

\fi

\usepackage{graphicx}
\usepackage{subfigure}
\usepackage{epstopdf}
\usepackage{amssymb}
\usepackage[cmex10]{amsmath}
\usepackage{stfloats}

\usepackage{subfigure}
\usepackage{tabularx}
\usepackage{verbatim}
\usepackage{url}
\usepackage{bm}
\usepackage{algorithm}
\usepackage{algorithmic}
\usepackage{stfloats}
\usepackage{cases}
\usepackage{cite}
\usepackage[square, comma, sort&compress, numbers]{natbib}
\usepackage{amsthm}

\newtheorem{remark}{Remark}
\begin{document}

\title{Cooperative Rate-Splitting for Secrecy Sum-Rate Enhancement in Multi-antenna Broadcast Channels}

\author{
	\vspace{0.4cm}
	\IEEEauthorblockN{Ping Li\IEEEauthorrefmark{1},
		Ming Chen\IEEEauthorrefmark{1},
		Yijie Mao\IEEEauthorrefmark{2},
		Zhaohui Yang\IEEEauthorrefmark{3},
		Bruno Clerckx\IEEEauthorrefmark{2},
		and Mohammad Shikh-Bahaei\IEEEauthorrefmark{3}}\\
	\IEEEauthorblockA{\IEEEauthorrefmark{1}National Mobile Communications Research Laboratory, Southeast University, China.}\\
	\IEEEauthorblockA{\IEEEauthorrefmark{2}Imperial College London, United Kingdom. }\\
	\IEEEauthorblockA{\IEEEauthorrefmark{3}Centre for Telecommunications Research, King's College London, United Kingdom.}\\
	\IEEEauthorblockA{Emails: \IEEEauthorrefmark{1}\{liping, chenming\}@seu.edu.cn, \IEEEauthorrefmark{2}\{y.mao16, b.clerckx\}@imperial.ac.uk, \IEEEauthorrefmark{3}\{yang.zhaohui, m.sbahaei\}@kcl.ac.uk}
	\vspace{-0.5cm}}

\maketitle

\begin{abstract}
In this paper, we employ Cooperative  Rate-Splitting (CRS) technique to enhance the Secrecy Sum Rate (SSR) for the Multiple Input Single Output (MISO) Broadcast Channel (BC), consisting of two legitimate users and one eavesdropper, with perfect Channel State Information (CSI) available at all nodes. For CRS based on the three-node relay channel, the transmitter splits and encodes the messages of legitimate users into common and private streams based on Rate-Splitting (RS). With the goal of maximizing SSR, the proposed CRS strategy opportunistically asks the relaying legitimate user to forward its decoded common message. During the transmission, the eavesdropper keeps wiretapping silently. To ensure secure transmission, the common message is used for the dual purpose, serving both as a desired message and Artificial Noise (AN) without consuming extra transmit power comparing to the conventional AN design. Taking into account the total power constraint and the Physical Layer (PHY) security, the precoders and time-slot allocation are jointly optimized by solving the non-convex SSR maximization problem based on Sequential Convex Approximation (SCA) algorithm. Numerical results show that the proposed CRS secure transmission scheme outperforms existing Multi-User Linear Precoding (MU-LP) and Cooperative Non-Orthogonal Multiple Access (C-NOMA) strategies. Therefore, CRS is a promising strategy to enhance the PHY security in multi-antenna BC systems. 
\end{abstract}

\begin{IEEEkeywords}
Cooperative rate-splitting, physical layer security, relay broadcast channel, secrecy sum rate, success convex approximation
\end{IEEEkeywords}

\IEEEpeerreviewmaketitle

\section{Introduction}
\IEEEPARstart{C}{ooperative}  relaying (CR), introduced in \cite{van1971three}, is an efficient technique to improve both the reliability and throughput of wireless networks. Recently, Cooperative Rate-Splitting (CRS), based on the three-node relay channel where the transmitter is equipped with multiple antennas, has been proposed as a more flexible transmission strategy than existing cooperative strategies and its non-cooperative counterpart \cite{zhang2019cooperative,mao2019max}. It achieves an explicit rate region improvement compared to the conventional baseline schemes in a wide range of propagation conditions. CRS is designed based on Rate-Splitting Multiple Access (RSMA), which is a promising multiple access technique that outperforms conventional Space Division Multiple Access (SDMA) and Non-Orthogonal Multiple Access (NOMA) \cite{clerckx2019rate,mao2018rate,mao2019rate,mao2018energy}. The core of RSMA framework is to split each transmitted message into a common and a private part, and pack all users' common parts on top of all the private parts. The combined common message is encoded using a codebook shared among all users and it can be decoded by all users, while the private message only can be decoded by the corresponding user. It is worth mentioning that each receiver decodes the common message first by regarding all private parts as interference, and then decodes the each private part after removing the common part via Successive Interference Cancellation (SIC). Such dynamic interference management strategy generalizes the two extreme schemes, i.e., treating all interference as noise and decoding all interference \cite{clerckx2016rate}. Hence, RSMA is capable to provide effective enhancement in spectral efficiency, energy efficiency, reliability and Channel State Information (CSI) feedback overhead  reduction \cite{clerckx2019rate,mao2018rate,clerckx2016rate,mao2019rate,mao2018energy,joudeh2016robust,dai2017multiuser,mao2019max}.   

Due to the broadcast characteristics of wireless communication, data transmission between devices is easily exposed to eavesdroppers, which poses a challenge to secure transmission. Instead of employing cryptographic techniques at the network layer, the Physical Layer (PHY) security enhances confidentiality by utilizing the reciprocity and randomness of the wireless channel. In \cite{wyner1975wire}, Wyner proposed the wiretap channel model and demonstrated that the legitimate user can demodulate correctly while ensuring the eavesdroper obtains nothing useful from the message, when the channel quality of the eavesdropper is worse than that of the legitimate receiver. The Wyner's discrete memoryless eavesdropping channel model is further extended to Broadcast Channels (BCs) \cite{csiszar1978broadcast} and Gaussian channels \cite{leung1978gaussian}, and it is proved in \cite{csiszar1978broadcast,leung1978gaussian} that the confidential capacity of the additive Gaussian channel is the difference between the capacity of the legitimate channel and that of the eavesdropping channel. Existing PHY security transmission technologies mainly include beamforming technology and Artificial Noise (AN) design \cite{zhao2015robust,zhang2016secure,mei2016robust,zhu2017beamforming,zheng2018secure}. In particular, the generation of AN needs to spare part of the transmit power to ensure secure transmission.

In this work, contrary to most of the existing works on PHY security which are based on Multi-User Linear Precoding (MU-LP) and NOMA \cite{zhao2015robust,zhang2016secure,mei2016robust,zhu2017beamforming,zheng2018secure}, we further explore the benefits of CRS for enhancing the PHY security of a multi-antenna BC. Specifically, we employ CRS technique to maximize the Secrecy Sum Rate (SSR) of Multiple Input Single Output (MISO) BC, consisting of a base station, two legitimate users and an eavesdropper. Aiming to maximize the SSR, CRS opportunistically asks the relaying legitimate user with strong CSI to forward its decoded common message to the user with weak CSI. During the whole transmission, the eavesdropper keeps wiretapping silently. It is worth mentioning that the common message is used for the dual purpose, serving as both a desired message and AN without consuming extra transmit power comparing with the common AN design adopted in the literature \cite{mei2016robust,zhu2017beamforming,zheng2018secure}.
To solve the non-convex SSR problem, we decompose the joint beamforming and time-slot allocation optimization problem into several sub-problems and convert them into a series of approximated convex problems based on the Sequential Convex Approximation (SCA) algorithm. Simulation results show that the proposed CRS secure transmission strategy achieves a higher system SSR performance compared to MU-LP and Cooperative NOMA (C-NOMA). Hence, CRS is an effective technique to improve the PHY security in multi-antenna BC. 

\section{System Model And Problem Formulation}
\subsection{System Model}
\begin{figure}[t]	
	\centering	\includegraphics[width=8cm]{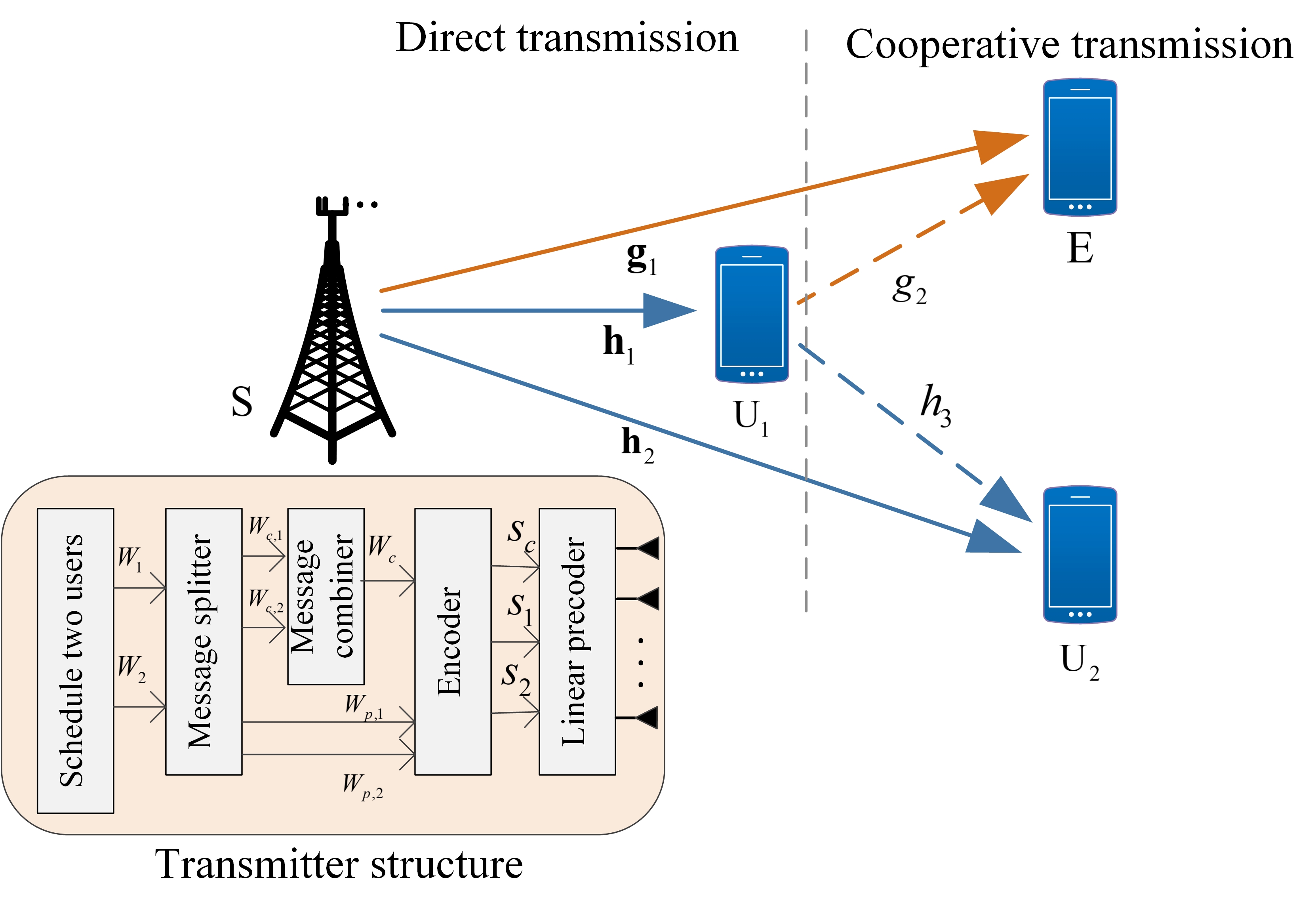}
	\vspace{-1em}
	\caption{Cooperative Rate-Splitting scheme. }	
	\label{fig1}
	\vspace{-0.5em}
\end{figure}

Consider a broadcast cooperative network as shown in Fig.~\ref{fig1}, which consists of a base station ($ S $) with $ N_T \left( N_T \geq 2 \right)  $ transmit antennas, two single-antenna legitimate users ($ U_1 $, $ U_2 $), and a potential single-antenna eavesdropper ($ E $). Without loss of generality, we assume that $ U_1 $ experiences a better channel than $ U_2 $\footnote{The norm of the CSI of $ U_1 $ is higher than that of $ U_2 $.}. Hence, we can regard $ U_1 $ as a potential relaying user to assist the signal processing of $ U_2 $ via the Non-regenerative Decode-and-Forward \cite{mesbah2008power} principle. The wireless channels at links $ S \rightarrow U_1 $, $ S \rightarrow U_2 $, $ S \rightarrow E $, $ U_1 \rightarrow U_2 $, $ U_1 \rightarrow E $ are expressed as $ {\bf{h}}_1 $, $ {\bf{h}}_2 $, $ {\bf{g}}_1 $, $ h_3 $, $ g_2 $, respectively. According to the Rate-Splitting (RS) principle, the message $ W_k $ intended to user $ k $ is split into one common part $ W_{c,k} $ and private part $ W_{p,k} $, $ k= 1,2 $. The common parts of both legitimate users are combined into one super common message $ W_c $ encoded by a shared codebook, while the private part of each user is processed individually. The resulting encoded three streams $ {\bf{s}} = {[{s_c},{s_1},{s_2}]^T} $ are precoded using ${\bf{P}} = [{{\bf{p}}_c},{{\bf{p}}_1},{{\bf{p}}_2}] \in {\mathbb{C}}^{{N_T} \times 3}$, and the superposed transmit signal is
\begin{equation}\label{expx}
{\bf{x}} = {\bf{Ps}} = {{\bf{p}}_c}{s_c} + {{\bf{p}}_1}{s_1} + {{\bf{p}}_2}{s_2},
\end{equation}
where $ {\bf{p}}_c \in {\mathbb{C}}^{{N_T} \times 1 }$ and $ {\bf{p}}_k \in {\mathbb{C}}^{{N_T} \times 1 }$, $ k=1,2 $ are the precoding vector for the common stream and the private stream transmitted for $ U_k $, respectively. Supposing that ${\mathbb{E}}\left[{\bf{s}}{{\bf{s}}^H}\right] = {{\bf{I}}} $, we have $ {\text{tr}}\left({\bf{P}}{{\bf{P}}^H}\right) \le {P_T} $ and $ P_T $ is the transmit power constraint at transmit node $ S $.

\begin{figure}[t]	
	\centering	\includegraphics[width=8cm]{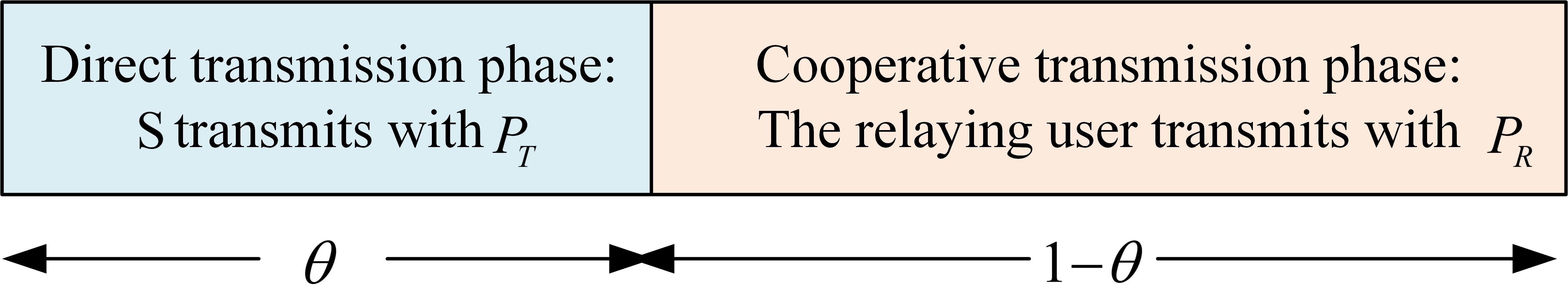}
	\vspace{-1em}
	\caption{Time-slot allocation for the two transmission phases.}	
	\label{fig2}
	\vspace{-1em}
\end{figure}

The relaying user is operating in half-duplex mode, i.e., two consecutive slots are required to complete the communication phases. As shown in Fig.~\ref{fig2}, $\theta {\kern 1pt} (0 \textless \theta  \le 1)$ is the fraction of time assigned to the direct transmissions ($ S \rightarrow U_1 $, $ S \rightarrow U_2 $, $ S \rightarrow E $), and the remaining is used for the cooperative transmissions ($ U_1 \rightarrow U_2 $, $ U_1 \rightarrow E $). The eavesdropper $ E $ wiretaps the signal transmitted to the legitimate users in both transmission phases. Different from the conventional half-duplex mode with equal-time slot allocation for the two hops, $\theta$ is dynamically adapted to the channel condition in this work. During the direct transmission phase (Phase $ \rm{\uppercase\expandafter{\romannumeral1}} $), $ S $ transmits information signal $ \bf{x} $ to the two legitimate users. The respective receive signal at $U_k$ and $E$ are 
\begin{equation} 
y_k^{(1)} = {\bf{h}}_k^H{\bf{x}} + {n_k},
\end{equation}
\begin{equation} 
y_e^{(1)} = {{\bf{g}}_1^H}{\bf{x}} + {n_{e1}},
\end{equation}
where the superscript ``$ (1) $'' represents Phase $ \rm{\uppercase\expandafter{\romannumeral1}} $, $ k=1,2 $. The terms $n_k\sim\mathcal {CN}(0,\sigma_k^2)$ and $n_{e1}\sim\mathcal {CN}(0,\sigma_{e_1}^2)$ are the Additive White Gaussian Noise (AWGN). In this paper, we assume  perfect CSI at the transmitter (CSIT) and the receiver (CSIR). 

Each legitimate user first decodes the common stream $ s_c $ by regarding both private streams as noise. Therefore, the Signal to Interference plus Noise Ratio (SINR) of decoding the common message $ s_c $ at $ U_k $ is
\begin{equation} 
\gamma _{c,k}^{(1)} = \frac{{{{\left| {{\bf{h}}_k^H{{\bf{p}}_c}} \right|}^2}}}{{{{\left| {{\bf{h}}_k^H{{\bf{p}}_1}} \right|}^2} + {{\left| {{\bf{h}}_k^H{{\bf{p}}_2}} \right|}^2} + \sigma _k^2}},\quad k =1,2.
\end{equation}

In the cooperative transmission phase (Phase $ \rm{\uppercase\expandafter{\romannumeral2}} $), the relaying user $ U_1 $ forwards the re-encoded common stream $ s_c $ by employing a different codebook from that of $ S $ to $ U_2 $ at a power level $ P_R $. The received signals at both $ U_2 $ and eavesdropper $ E $ are written respectively as
\begin{equation}\label{exp_5}
y_2^{(2)} = {h_3^H}\sqrt {{P_R}} {s_c} + {n_3},
\end{equation} 
\begin{equation}
y_e^{(2)} = {g_2^H}\sqrt {{P_R}} {s_c} + {n_{e2}}.
\end{equation}
The superscript ``$ (2) $'' represents Phase $ \rm{\uppercase\expandafter{\romannumeral2}} $.  $n_3\sim\mathcal {CN}(0,\sigma_3^2)$, $n_{e2}\sim\mathcal {CN}(0,\sigma_{e2}^2)$ are the AWGN at the nodes of $ U_2 $ and $ E $ in Phase $ \rm{\uppercase\expandafter{\romannumeral2}} $, respectively. Based on (\ref{exp_5}), the SINR of $ U_2 $ is expressed as
\begin{equation}
\gamma _{c,2}^{(2)} = \frac{{{{\left| {h_3^H\sqrt {{P_R}} } \right|}^2}}}{{\sigma _3^2}}.
\end{equation}

To ensure that the common message is decodable at the two legitimate users, the corresponding achievable rate of decoding the common message should be
\begin{equation}\label{expRc}
{R_c} = \mathop{\min} \{ {R_{c,1}},{R_{c,2}}\},
\end{equation} 
where ${R_{c,1}} = \theta {\log _2}(1 + \gamma _{c,1}^{(1)})$ is the achievable rate of decoding the common stream $s_c$ at $U_1$, and ${R_{c,2}} = \theta {\log _2}(1 + \gamma _{c,2}^{(1)}) + (1 - \theta ){\log _2}(1 + \gamma _{c,2}^{(2)})$ is the corresponding achievable rate at $ U_2 $.

The legitimate users decode the private messages individually after removing the common part via SIC. Hence, the corresponding SINR and the corresponding achievable rate of decoding $ s_k $ at $ U_k $ are respectively written as
\begin{equation} 
\gamma _{p,k}^{(1)} = \frac{{{{\left| {{\bf{h}}_k^H{{\bf{p}}_k}} \right|}^2}}}{{{{\left| {{\bf{h}}_k^H{{\bf{p}}_i}} \right|}^2} + \sigma _k^2}},\quad i \ne k,
\end{equation}
\begin{equation}
{R_{p,k}} = \theta {\log _2}(1 + \gamma _{p,k}^{(1)}),
\end{equation}
where $ i,k=1,2 $.

As the precoder of the common stream designed at the BS only ensures the decodability of the common stream at the two legitimate users, the eavesdropper $E$ may not be able to decode $s_c$. Hence, the common message can be treated as AN at $ E $. To meet this requirement, the condition, i.e., $ {C_{c,e}} \le {R_c} $, should be satisﬁed, where $ {C_{c,e}} $ represents the sum rate of the common message from $ S $ to $ E $ and from $ U_1 $ to $ E $. The corresponding SINR of private message at $ E $ is degraded by the existence of the common message. The receiving SINRs of decoding the common streams at each phase and the private stream at $ E $ are written as
\begin{equation}
\gamma _{c,e}^{(1)} = \frac{{{{\left| {{{\bf{g}}_1^H}{{\bf{p}}_c}} \right|}^2}}}{{{{\left| {{{\bf{g}}_1^H}{{\bf{p}}_1}} \right|}^2} + {{\left| {{{\bf{g}}_1^H}{{\bf{p}}_2}} \right|}^2} + \sigma _{e1}^2}},
\end{equation}
\begin{equation}
\gamma _{c,e}^{(2)} = \frac{{{{\left| {g_2^H\sqrt {{P_R}} } \right|}^2}}}{{\sigma _{e2}^2}},
\end{equation}
\begin{equation}
\gamma _{k,e}^{(1)} = \frac{{{{\left| {{{\bf{g}}_1^H}{{\bf{p}}_k}} \right|}^2}}}{{{{\left| {{{\bf{g}}_1^H}{{\bf{p}}_c}} \right|}^2} + {{\left| {{{\bf{g}}_1^H}{{\bf{p}}_j}} \right|}^2} + \sigma _{e1}^2}},\quad j \ne k,
\end{equation}
respectively, where $ j,k=1,2 $. The corresponding achievable rate of decoding $s_c$ and $s_k$ at $ E $ are expressed as
\begin{equation}
{C_{c,e}} = \theta {\log _2}(1 + \gamma _{c,e}^{(1)}) + (1 - \theta ){\log _2}(1 + \gamma _{c,e}^{(2)}),
\end{equation}
\begin{equation}
{C_{k,e}} = \theta {\log _2}(1 + \gamma _{k,e}^{(1)}),{\kern 1pt}k =1,2.
\end{equation}
Thus, the achievable SSR from $ S $ to $ U_1 $ and $ U_2 $ is given as
\begin{equation}
R^{sec}_{tot} = R_c^{\sec } + R^{sec}_{p,1}+R^{sec}_{p,2},
\end{equation} 
where $R_c^{\sec } = {[{R_c} - {C_{c,e}}]^ + }$, $R_{p,k}^{\sec } = {[{R_{p,k}} - {C_{k,e}}]^ + }$, $ k=1,2 $, represent the achievable secrecy rate of the common message and the private message transmitted to $ U_k $, respectively. The operation $ [x]^+=\max\{x,0\} $.

\subsection{Problem Formulation}
In this work, we aim to optimize the precoding $\mathbf{P}$ and time-slot allocation $\theta$ with the objective of maximizing the SSR subject to transmit power constraint. Mathematically, the SSR optimization problem is written as
\begin{subequations}\label{sysmax1}
	\setlength\abovedisplayskip{8pt}
	\setlength\belowdisplayskip{8pt}
	\begin{align}
	\mathop{\max}\limits_{{\bf{P}},\theta}\quad&R^{sec}_{tot}\\
	\textrm{s.t.} \quad\:
	&C_{c,e} \le R_{c},\\
	&\text{tr}\left({\bf{P}}{{\bf{P}}^H}\right) \le {P_T},0 \textless \theta  \le 1,
	\end{align}
\end{subequations}
where $ P_T $ is the maximum transmit power at $ S $. The constraint (\ref{sysmax1}b) ensures the eavesdropper $ E $ 
is unable to decode the legitimate common message, while (\ref{sysmax1}c) presents the transmit power constraint and gives the value range of $ \theta $.
\section{Optimization Solution}
Before solving the optimization problem \eqref{sysmax1}, we first analyze its convexity. The $ \mathop{\max} $ operator is convex, while the $ \mathop{\max} \max $ is non-convex. Furthermore, the elements in max functions, i.e. $ {R_c} - {C_{c,e}} $ and $ {R_{p,k}} - {C_{k,e}} $, are neither convex nor concave. To simplify the original optimization problem, we first decompose \eqref{sysmax1}
to draw the elments out of the max operators below, then we perform a series of linearizations to deal with its non-convexity. We identify four different cases:  
\begin{subequations}\label{sysmax_1}\vspace{-1em}
\setlength\belowdisplayskip{0.5pt}
	\begin{align}
	Case {\kern 1pt} 1: \mathop{\max}\limits_{{\bf{P}},\theta}\quad&{R_c} - {C_{c,e}}+{R_{p,1}} - {C_{1,e}}+{R_{p,2}} - {C_{2,e}}\\
	\textrm{s.t.} \quad\:
	&{R_{p,1}} \ge {C_{1,e}},{R_{p,2}} \ge {C_{2,e}},\\
	&(\ref{sysmax1}b),(\ref{sysmax1}c),
	\end{align}
\end{subequations}
\begin{subequations}\label{sysmax_2}
\setlength\abovedisplayskip{0.5pt}
\setlength\belowdisplayskip{0.5pt}
	\begin{align}
	Case {\kern 1pt} 2:
	\mathop{\max}\limits_{{\bf{P}},\theta}\quad&{R_c} - {C_{c,e}}+{R_{p,1}} - {C_{1,e}}\\
	\textrm{s.t.} \quad\:
	&{R_{p,1}} \ge {C_{1,e}},{R_{p,2}} \le {C_{2,e}},\\
	&(\ref{sysmax1}b),(\ref{sysmax1}c),
	\end{align}
\end{subequations}
\begin{subequations}\label{sysmax_3}
\setlength\abovedisplayskip{0.5pt}
\setlength\belowdisplayskip{0.5pt}
	\begin{align}
	Case {\kern 1pt} 3:
	\mathop{\max}\limits_{{\bf{P}},\theta}\quad&{R_c} - {C_{c,e}}+{R_{p,2}} - {C_{2,e}}\\
	\textrm{s.t.} \quad\:
	&{R_{p,1}} \le {C_{1,e}},{R_{p,2}} \ge {C_{2,e}},\\
	&(\ref{sysmax1}b),(\ref{sysmax1}c),
	\end{align}
\end{subequations}
\begin{subequations}\label{sysmax_4}
	\vspace{-1.5em}

	\begin{align}
	Case {\kern 1pt} 4:
	\mathop{\max}\limits_{{\bf{P}},\theta}\quad& {R_c} - {C_{c,e}}\\
	\textrm{s.t.} \quad\:
	&{R_{p,1}} \le {C_{1,e}},{R_{p,2}} \le {C_{2,e}},\\
	&(\ref{sysmax1}b),(\ref{sysmax1}c).
	\end{align}	
\end{subequations}

As the non-convexity among all cases are similar, we next specify the optimization framework to solve Problem \eqref{sysmax_1} for simplicity. Note that the method for solving problem \eqref{sysmax_1} can be easily applied to problems \eqref{sysmax_2} - \eqref{sysmax_4}. The solution to problem \eqref{sysmax1} lies in the solution to one problem in \eqref{sysmax_1}-\eqref{sysmax_4} with the highest objective value.
  
To handle the non-convex problem \eqref{sysmax_1}, we first equivalently rewrite \eqref{sysmax_1} as 
\begin{subequations}\label{sysmax2}
	\begin{align}
	\mathop{\max}\limits_{\bf{P},\theta,{\boldsymbol{\alpha}}_{p},{\boldsymbol{\alpha}}_{c}} & \mathop{\min} \{ \alpha_{c,1},\alpha_{c,2}\} +\sum\nolimits_{k} \left(\alpha_{p,k}-\alpha_{k,e}\right)-\alpha_{c,e}\\
	\textrm{s.t.} \quad\:
	&R_{p,k} \ge \alpha_{p,k},\\
	&R_{c,k} \ge \alpha_{c,k},\\
	&C_{k,e} \le \alpha_{k,e},\\
	&C_{c,e} \le \alpha_{c,e},\\
	&{\alpha_{p,1}} \ge {\alpha_{1,e}},{\alpha_{p,2}} \ge {\alpha_{2,e}},\\
	&\alpha_{c,e} \le \alpha_{c,1},\alpha_{c,e} \le \alpha_{c,2},\\  
	&(\ref{sysmax1}c),
	\end{align}
\end{subequations}
with auxiliary variable vectors $ \boldsymbol{\alpha}_{p}=[ \alpha_{p,1},\alpha_{p,2},\alpha_{1,e},\alpha_{2,e} ] $ and $ \boldsymbol{\alpha}_{c}=[ \alpha_{c,1},\alpha_{c,2},\alpha_{c,e}] $ which are introduced to represent the corresponding rates of decoding the common and private streams at legitimate users $U_1, U_2$ and  eavesdropper $ E $, $ k=1,2 $. The difficulty in solving \eqref{sysmax2} is due to the constraints (\ref{sysmax2}b)-(\ref{sysmax2}e). To further expose the hidden convexity of these inequalities, in the derivation below, we introduce several vectors $ \boldsymbol{\beta}_{p}=[ \beta_{p,1},\beta_{p,2},\beta_{1,e},\beta_{2,e} ] $, $ \boldsymbol{\beta}_{c}=[ \beta_{c,1},\beta_{c,2},\beta_{c,e}] $, $ \boldsymbol{\rho}_{p}=[ \rho_{p,1},\rho_{p,2},\rho_{1,e} ,\rho_{2,e}] $, $ \boldsymbol{\rho}_{c}=[ \rho_{c,1},\rho_{c,2},\rho_{c,e} ] $, where $ \beta_{p,k} $, $ \beta_{c,k} $, $ \rho_{p,k} $, $ \rho_{c,k} $ respectively denote the achievable rates (without $\theta$) and the SINR of private and common streams in Phase $ \rm{\uppercase\expandafter{\romannumeral1}} $ at $ U_k $, while $ \beta_{k,e} $, $ \beta_{c,e} $, $ \rho_{k,e} $, $ \rho_{c,e} $ represent the achievable rate (without $\theta$) and the SINR of decoding the private and the common streams in Phase $ \rm{\uppercase\expandafter{\romannumeral1}} $ at $ E $, $ k=1,2 $. To handle (\ref{sysmax2}b), we rewrite it as
\begin{subequations}\label{expre1}
	\begin{align}
	&\theta \log_2 \left( 1+\rho_{p,k} \right)  \ge \alpha_{p,k},\\
	&\gamma_{p,k}^{(1)} \ge \rho_{p,k}.
	\end{align}
\end{subequations}
Note that (\ref{expre1}a) is equivalently replaced by constraints
\begin{subequations}\label{expre2}
	\begin{align}
	&\theta \beta_{p,k}  \ge \alpha_{p,k},\\
	&\log_2 \left( 1+\rho_{p,k} \right)  \ge \beta_{p,k} \Leftrightarrow 1+\rho_{p,k}-2^{\beta_{p,k}} \ge 0.
	\end{align}
\end{subequations}
Based on \eqref{expre1} and \eqref{expre2}, (\ref{sysmax2}b) becomes
\begin{subnumcases}
{(\ref{sysmax2}b) \Leftrightarrow} 
\theta \beta_{p,k}  \ge \alpha_{p,k}, \label{exp18ba} \\
\gamma_{p,k}^{(1)} \ge \rho_{p,k},\label{exp18bb}\\
1+\rho_{p,k}-2^{\beta_{p,k}} \ge 0. \label{exp18bc}
\end{subnumcases}
To deal with the non-convex constraints (\ref{exp18ba}) and (\ref{exp18bb}), we adopt the following approximation. For constraint (\ref{exp18ba}), $ \theta \beta_{p,k} $ is equivalent to $ \theta \beta_{p,k} =\frac{1}{4}( \theta + \beta_{p,k} )^2- \frac{1}{4}( \theta - \beta_{p,k} )^2 $. Hence, approximated at the point $ ( \theta^{[n]}, \beta_{p,k}^{[n]})  $ by the first-order Taylor expansion of $ ( \theta + \beta_{p,k} )^2 $, $ \theta \beta_{p,k} $ is given by
\begin{multline}
\theta \beta_{p,k}  \ge \frac{1}{2} \left( \theta^{[n]} + \beta_{p,k}^{[n]} \right)\left( \theta+\beta_{p,k}\right)-\frac{1}{4}\left( \theta^{[n]} + \beta_{p,k}^{[n]} \right)^2 -\\ \frac{1}{4}\left( \theta-\beta_{p,k}\right)^2 \buildrel \Delta \over = \Phi^{[n]}(\theta,\beta_{p,k}),
\end{multline}
where the superscript ``$[n]$'' denotes the $n$-th iteration. Therefore, (\ref{exp18ba}) is rewritten as 
\begin{equation}\label{exp18baa}
\Phi^{[n]}(\theta,\beta_{p,k}) \ge \alpha_{p,k},
\end{equation}
where $ k=1,2 $. As for (\ref{exp18bb}), we convert it to Difference-of-Convex (DC) form:
\begin{equation}\label{eqn3}
{\left| {{\bf{h}}_k^H{{\bf{p}}_i}} \right|^2} + \sigma _k^2 - \frac{{{{\left| {{\bf{h}}_k^H{{\bf{p}}_k}} \right|}^2}}}{{{\rho _{p,k}}}}  \le 0,i \ne k,
\end{equation}
where $ i,k=1,2 $. To deal with \eqref{eqn3}, we relax concave parts of the DC constraints with their ﬁrst-order Taylor expansions. (\ref{eqn3}) is approximated around the point $ ( {\bf{p}}_k^{[n]}, {\rho}_{p,k}^{[n]} )  $ at iteration $ n $ by 
\begin{equation}\label{eqn4}
{\left| {{\bf{h}}_k^H{{\bf{p}}_i}} \right|^2} + \sigma _k^2 - {\Psi ^{[n]}}({{\bf{p}}_k},{{\bf{h}}_k},{{\bf{\rho }}_{p,k}}) \le 0,i \ne k,
\end{equation}
where $ {\Psi ^{[n]}}({{\bf{p}}_k},{{\bf{h}}_k},{{\rho }}_{p,k}) = \frac{{2\Re\{ {{({\bf{p}}_k^{[n]})}^H}{{\bf{h}}_k}{\bf{h}}_k^H{{\bf{p}}_k}\} }}{{\rho _{p,k}^{[n]}}} - \frac{{{{\left| {{\bf{h}}_k^H{\bf{p}}_k^{[n]}} \right|}^2}{\rho _{p,k}}}}{{{{(\rho _{p,k}^{[n]})}^2}}} $ and $ i,k=1,2 $. We also rewrite (\ref{sysmax2}c) as
\begin{subequations}\label{expre3}
\begin{align}
&\Phi^{[n]}(\theta,\beta_{c,1}) \ge \alpha_{c,1}, \\
&\Phi^{[n]}(\theta,\beta_{c,2}) \ge \alpha_{c,2}-R_{c,2}^{(2)}, \\
&\sum\nolimits_{m}{\left| {{\bf{h}}_k^H{{\bf{p}}_m}} \right|^2} + \sigma _k^2 - {\Psi ^{[n]}}({{\bf{p}}_c},{{\bf{h}}_k},{{{\rho }}_{c,k}}) \le 0,\\
&1+\rho_{c,k}-2^{\beta_{c,k}} \ge 0,
\end{align}
\end{subequations}
where $ m,k=1,2 $, $ R_{c,2}^{(2)}=(1-\theta) \log_2 ( 1+\gamma_{c,2}^{(2)}) $. Following the same approach, we rewrite (\ref{sysmax2}d) as
\begin{subnumcases}
{(\ref{sysmax2}d) \Leftrightarrow} 
\theta \beta_{k,e}  \le \alpha_{k,e},\label{exp18da} \\
\gamma_{k,e}^{(1)} \le \rho_{k,e},\label{exp18db}\\
1+\rho_{k,e}-2^{\beta_{k,e}} \le 0.\label{exp18dc}
\end{subnumcases}
The differences in processing methods between (\ref{sysmax2}b) and (\ref{sysmax2}d) are that
\begin{itemize}
	\item [1)] 
	for (\ref{exp18da}), we expand $ ( \theta - \beta_{k,e} )^2  $ around $ ( \theta^{[n]}, {\beta}_{k,e}^{[n]})  $ to get the lower bound of $ \theta \beta_{k,e} $.       
	\item [2)]
	for (\ref{exp18db}), we transform it to DC form, i.e., $ {| {{\bf{g}}_1^H{{\bf{p}}_c}} |^2} + {| {{\bf{g}}_1^H{{\bf{p}}_j}} |^2} +
	\sigma _{e1}^2 - \frac{{{{| {{\bf{g}}_1^H{{\bf{p}}_k}} |}^2}}}{{{\rho _{k,e}}}} \ge 0,j,k=1,2,j \ne k  $, and we approximate $ {| {{\bf{g}}_1^H{{\bf{p}}_c}} |^2} + {| {{\bf{g}}_1^H{{\bf{p}}_j}} |^2} $ using its first-order Taylor expansion around the point $ ( {\bf{P}}^{[n]})  $. 
	\item [3)]
	for (\ref{exp18dc}), we approximate $ 2^{\beta_{k,e}} $ around the point $ (\beta_{k,e}^{[n]}) $.
\end{itemize}
Hence, (\ref{sysmax2}d) is expressed as
\begin{subequations}\label{eqn7}
\begin{align}
&\Theta^{[n]}(\theta,\beta_{k,e}) \le \alpha_{k,e}, \\
&{\Omega ^{[n]}({{\bf{p}}_c},{{\bf{p}}_j},{{\bf{g}}_1})}+ \sigma _{e1}^2-\frac{{{{\left| {{\bf{g}}_1^H{{\bf{p}}_k}} \right|}^2}}}{{{\rho _{k,e}}}}  \ge 0, j\ne k,\\
&1+\rho_{k,e}-{\Gamma}^{[n]} (\beta_{k,e}) \le 0,
\end{align}
\end{subequations}
where 
\begin{multline}\nonumber
\Theta^{[n]}(\theta,\beta_{k,e}) \buildrel \Delta \over = \frac{1}{4}( \theta + \beta_{k,e} )^2+\frac{1}{4}( \theta^{[n]}-\beta_{k,e}^{[n]})^2-\\\frac{1}{2} ( \theta^{[n]} - \beta_{k,e}^{[n]} )( \theta-\beta_{k,e}),
\end{multline}
\begin{multline}\nonumber
\Omega ^{[n]}({{{\bf{p}}_c}},{{{\bf{p}}_j}},{{\bf{g}}_1}) \buildrel \Delta \over = {2\Re\{ {{({\bf{p}}_c^{[n]})}^H}{{\bf{g}}_1}{\bf{g}}_1^H{{\bf{p}}_c}+{{({\bf{p}}_j^{[n]})}^H}{{\bf{g}}_1}{\bf{g}}_1^H{{\bf{p}}_j}\} }\\-{{| {{\bf{g}}_1^H{\bf{p}}_c^{[n]}} |}^2}-{{| {{\bf{g}}_1^H{\bf{p}}_j^{[n]}} |}^2},
\end{multline}
\begin{multline}\nonumber
 \Gamma^{[n]}(\beta_{k,e}) \buildrel \Delta \over = 2^{\beta_{k,e}^{[n]}} \left[1+(\ln{2})(\beta_{k,e}-\beta_{k,e}^{[n]})\right], 
\end{multline}
and $ j,k=1,2 $. Following the approximation of (\ref{sysmax2}d), we convert (\ref{sysmax2}e) as
\begin{subequations}\label{eqn8}
\begin{align}
&\Theta^{[n]}(\theta,\beta_{c,e}) \le \alpha_{c,e}-C_{c,e}^{(2)}, \\
&{\Omega ^{[n]}({{\bf{p}}_1},{{\bf{p}}_2},{{\bf{g}}_1})}+ \sigma _{e1}^2-\frac{{{{\left| {{\bf{g}}_1^H{{\bf{p}}_c}} \right|}^2}}}{{{\rho _{c,e}}}} \ge 0,\\
&1+\rho_{c,e}-{\Gamma}^{[n]} (\beta_{c,e}) \le 0.
\end{align}
\end{subequations}

Therefore, at iteration $ n $, based on the optimal solution $ ( \theta^{[n]},{\bf{P}}^{[n]},{\boldsymbol{\beta}}_{p}^{[n]},{\boldsymbol{\beta}}_{c}^{[n]},{\boldsymbol{\rho}}_{p}^{[n]},{\boldsymbol{\rho}}_{c}^{[n]}) $ obtained from the previous iteration $ n-1 $, we solve the following subproblem:
\begin{subequations}\label{sysmax4}
	\begin{align}
	\mathop{\max}\limits_{\bf{P},\theta,{\bf{\alpha}}_{p},{\bf{\alpha}}_{c},{\bf{\beta}}_{p} \atop {\bf{\beta}}_{c},{\bf{\rho}}_{p},{\bf{\rho}}_{c}} & \mathop{\min} \{ \alpha_{c,1},\alpha_{c,2}\} +\sum\nolimits_{k} \left(\alpha_{p,k}-\alpha_{k,e}\right)-\alpha_{c,e}\\
	\textrm{s.t.} \quad\:
	&(\ref{exp18bc}),(\ref{exp18baa}), (\ref{eqn4}),\\
	&(\ref{expre3}), (\ref{eqn7}),(\ref{eqn8}),(\ref{sysmax2}f),(\ref{sysmax2}g),(\ref{sysmax1}c).
	\end{align}
\end{subequations}

The proposed SCA-based algorithm is summarized in Algorithm~\ref{algorithm1}.
\begin{algorithm*}[h]
	\caption{Proposed SCA-based algorithm for solving problem \eqref{sysmax_1}}
	\begin{algorithmic}[1]\label{algorithm1}
		\STATE Initialize $n \leftarrow 0, t^{[n]} \leftarrow 0 , {\bf{P}}^{[n]},\theta^{[n]},{\boldsymbol{\beta}}_{p}^{[n]},{\boldsymbol{\beta}}_{c}^{[n]},{\boldsymbol{\rho}}_{p}^{[n]},{\boldsymbol{\rho}}_{c}^{[n]} $, set the value of $ \epsilon $;
		\REPEAT
		\STATE $n \leftarrow n+1 $;
		\STATE Solve problem \eqref{sysmax4} using $ {\bf{P}}^{[n-1]},\theta^{[n-1]},{\boldsymbol{\beta}}_{p}^{[n-1]},{\boldsymbol{\beta}}_{c}^{[n-1]},{\boldsymbol{\rho}}_{p}^{[n-1]},{\boldsymbol{\rho}}_{c}^{[n-1]} $, denote the optimal value of the objective function as $ t^{*} $ and the optimal solutions as $ {\bf{P}}^{*},\theta^{*},{\boldsymbol{\beta}}_{p}^{*},{\boldsymbol{\beta}}_{c}^{*},{\boldsymbol{\rho}}_{p}^{*},{\boldsymbol{\rho}}_{c}^{*} $; 
		\STATE Update $ t^{[n]} \leftarrow t^{*},{\bf{P}}^{[n]} \leftarrow {\bf{P}}^{*},\theta^{[n]} \leftarrow \theta^{*},{\boldsymbol{\beta}}_{p}^{[n]} \leftarrow {\boldsymbol{\beta}}_{p}^{*},{\boldsymbol{\beta}}_{c}^{[n]} \leftarrow {\boldsymbol{\beta}}_{c}^{*},{\boldsymbol{\rho}}_{p}^{[n]} \leftarrow {\boldsymbol{\rho}}_{p}^{*},{\boldsymbol{\rho}}_{c}^{[n]}\leftarrow {\boldsymbol{\rho}}_{c}^{*}$;
		\UNTIL $ |t^{[n]}-t^{[n-1]}| \leq \epsilon $;
	\end{algorithmic}
\end{algorithm*}

\begin{remark}
	\label{remark1}
	The SSR of the proposed CRS secure transmission is always larger than or equal to that of MU-LP and C-NOMA. In our paper, Non-cooperative Rate-Splitting (NRS) without user relaying is a special case of the proposed CRS when $\theta$ is fixed to 1.  As discussed in \cite{zhang2019cooperative}, NRS boils down to MU-LP when the power allocated to the common stream is 0, and CRS is more general than NRS and MU-LP. Moreover, CRS reduces to C-NOMA when the common stream encodes the entire message of one of the two users. Therefore, the SSR achieved by the proposed CRS is always larger than or equal to that of MU-LP and C-NOMA.
\end{remark}

\vspace{-1.5em}
\subsection{Convergence Analysis}
\vspace{-0.2em}
Following \cite{beck2010sequential}, we briefly demonstrate the convergence of Algorithm~\ref{algorithm1}. Let $ t^{[n]}$ denote the optimal objective which equals to $ \mathop{\min} \{ \alpha_{c,1}^{[n]},\alpha_{c,2}^{[n]}\} +\sum\nolimits_{k} (\alpha_{p,k}^{[n]}-\alpha_{k,e}^{[n]})-\alpha_{c,e}^{[n]} $ in our paper. Given a feasible initial point, due to the linear expansion in processing the original optimization problem, the solution generated by solving \eqref{sysmax4} at the iteration $ n-1 $ is the feasible point of \eqref{sysmax4} at iteration $ n $. Thus, the SCA-based algorithm can yields a non-decreasing sequence, i.e., $ t^{[n]} \geq t^{[n-1]} $. Due to the power constraint in (\ref{sysmax1}c), the series $ \{t^{[n]}\}_{n=1}^{n=\infty} $ is bounded above, hence, Algorithm~\ref{algorithm1} guarantees the convergence of the optimization problem \eqref{sysmax_1}.
\vspace{-1.0em} 
\subsection{Complexity Analysis}
\vspace{-0.2em}
Due to the exponential cone constraints (\ref{exp18bc}) and (\ref{expre3}d), problem \eqref{sysmax4} is a generalized nonlinear convex problem. An alternative efficient method is to
approximate (\ref{exp18bc}) and (\ref{expre3}d) through a sequence of Second Order Cone (SOC) via the successive approximation method \cite{ben2001polyhedral,tervo2015optimal}, and the SOC Programming (SOCP) can be solved via interior-point methods with complexity $ \mathcal{O} ([K{N}_T]^{3.5}) $, where $ K $ is the number of legitimate users. The total number of iterations required for convergence is approximated as $ \mathcal{O} (\log(\epsilon^{-1})) $, where $ \epsilon $ is the accuracy of the proposed Algorithm~\ref{algorithm1}. Therefore, considering the worst case, the computational complexity is $ \mathcal{O} (\log(\epsilon^{-1})) [K{N}_T]^{3.5} $.
\vspace{-1em}
\subsection{Generation of Initial Points}
\vspace{-0.2em}
From the convergence analysis above, it is obvious that if the initial solution $ ( {\bf{P}}^{[0]},\theta^{[0]},\boldsymbol{\beta}_p^{[0]},\boldsymbol{\beta}_c^{[0]},\boldsymbol{\rho}_p^{[0]},\boldsymbol{\rho}_c^{[0]}) $ are feasible to \eqref{sysmax4}, then the problems of the subsequent iterations are also feasible and solvable. Hence, we can solve a simple problem: $ \text{find}\{{\bf{P}},\theta|(\ref{sysmax_1}b),(\ref{sysmax_1}c)\} $, and denote the obtained solutions as $({\bf{P}}^{[0]},\theta^{[0]})  $, while the initial value of the remaining variables can be obtained by replacing the related inequality of \eqref{sysmax4} by equalities.
\section{Simulation Results}
In this section, we evaluate the performance of proposed algorithm. Following \cite{mao2019max}, we consider all channels are independent and identically distribute (i.i.d) complex Gaussian random entries. The variance of AWGN is $ 1 $. The links from $ S $ to legitimate users, $ U_1 $, $ U_2 $, and the eavesdropper $ E $ follow ${\bf{h}}_k\sim\mathcal {CN}(0,\sigma_{h_k}^2)$, ${\bf{g}}_1\sim\mathcal {CN}(0,\sigma_{g_1}^2)$, respectively. Supposing the channel gain of $ U_1 $ is stronger than that of $ U_2 $, the links from the relay $ U_1 $ to $ U_2 $ and $ E $ follow $h_3\sim\mathcal {CN}(0,\sigma_{h_3}^2)$, $g_1\sim\mathcal {CN}(0,\sigma_{g_2}^2)$, respectively, where $ k=1,2 $. In the following simulation, we assume $ \sigma_{g_1}^2=1, \sigma_{h_3}^2=1,\sigma_{g_2}^2=1$. The tolerance of the proposed SCA-based algorithm is set to $ \epsilon=10^{-3} $. Without loss of generality, we assume the transmit power at $ S $ and the relaying user are equal, i.e., $ P_T=P_R $.
\begin{figure}[t]
	\centering
	\subfigure[$ N_T=2,\sigma_{h_1}^2=1,\sigma_{h_2}^2=1 $] { \label{fig:a} 
	\includegraphics[width=8cm,height=5cm]{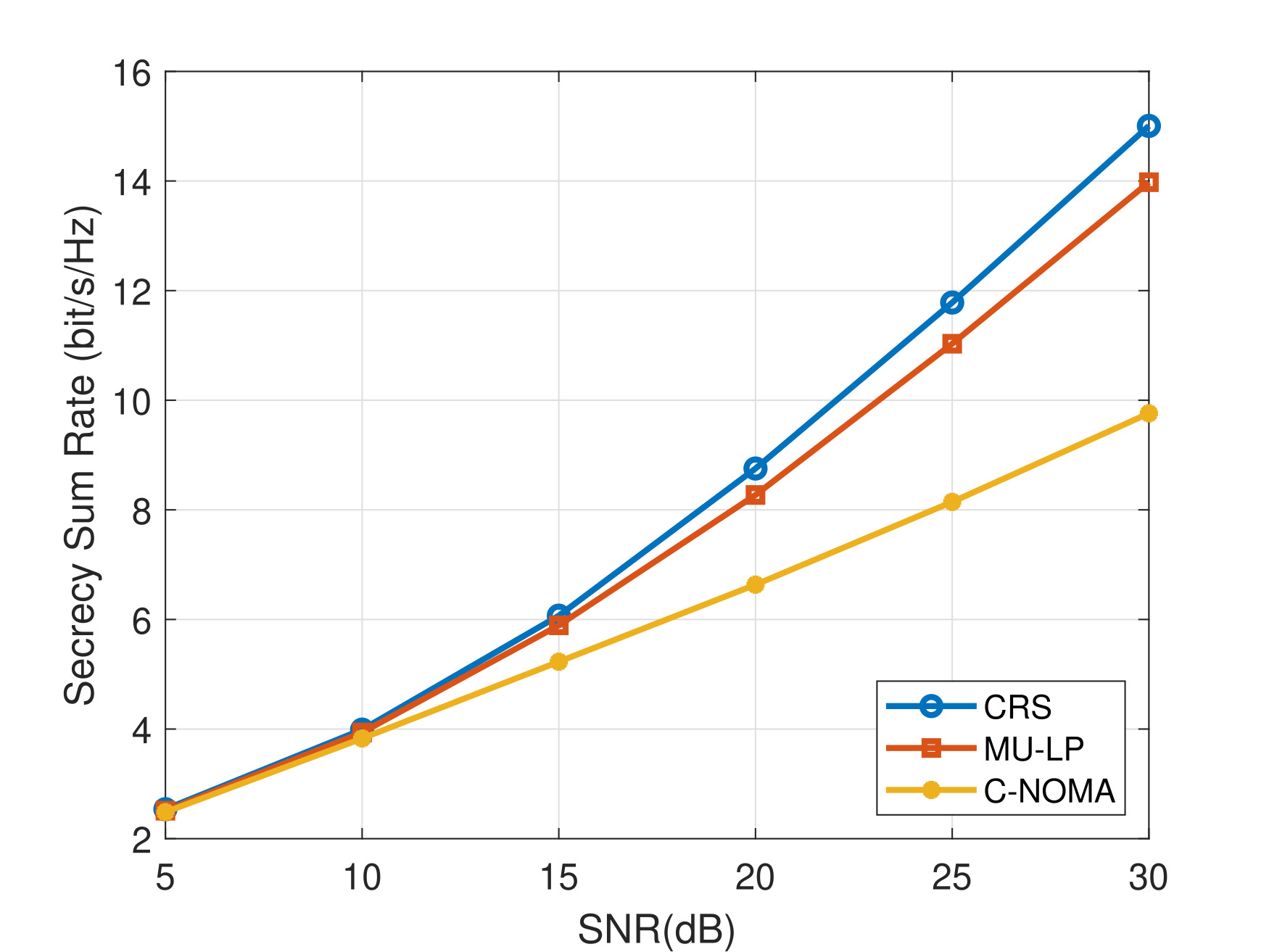}}
	\setlength\belowcaptionskip{-0.5pt}
	\subfigure[$ N_T=2,\sigma_{h_1}^2=1,\sigma_{h_2}^2=0.3 $] { \label{fig:b}
	\includegraphics[width=8cm,height=5cm]{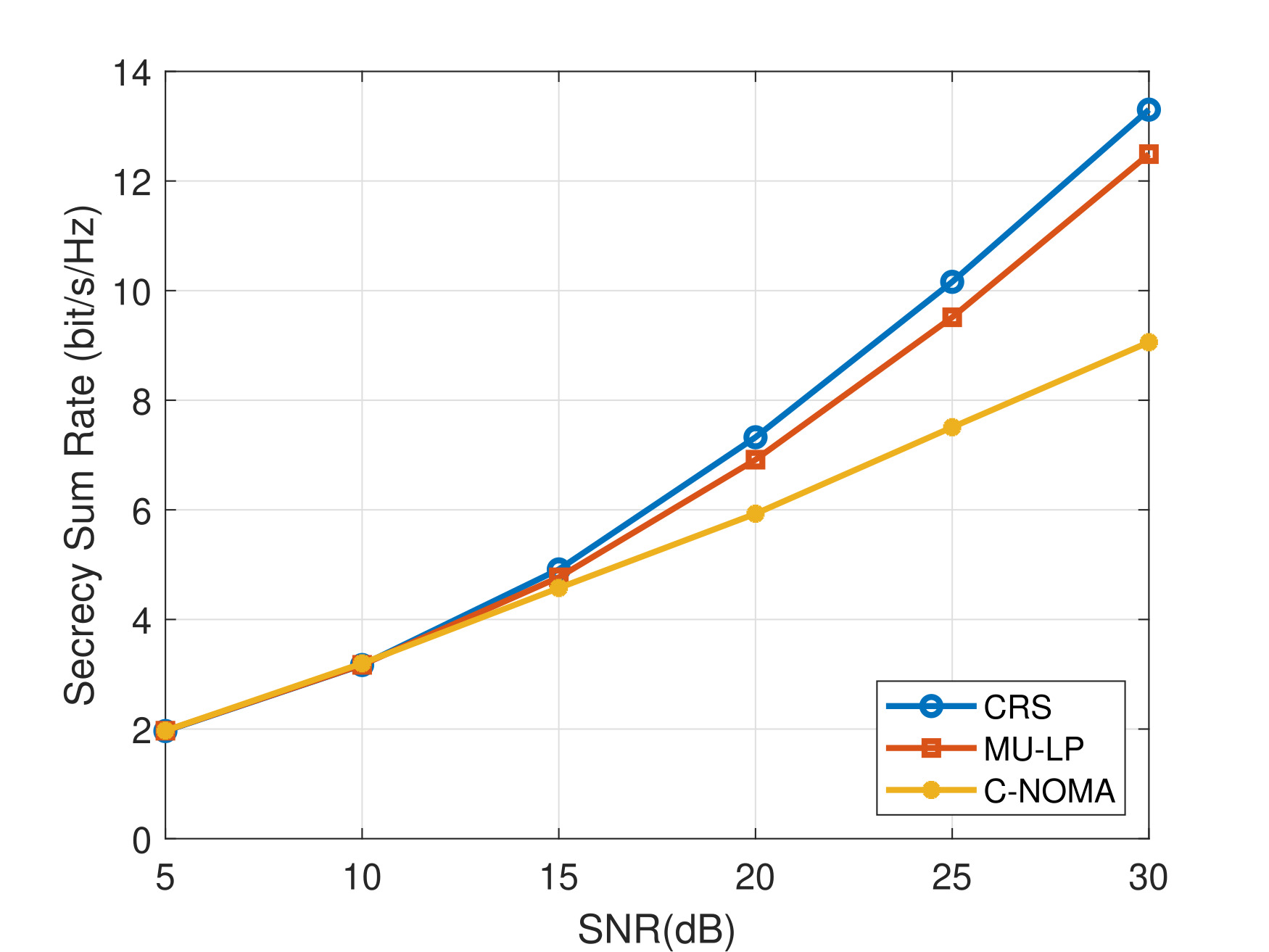}}
	\vspace{-0.5em}
	\caption{The average SSR versus SNR comparison of different strategies over 100 random channels, $ N_T=2 $.} 
	\label{fig3}
	\vspace{-1.5em}
\end{figure}

Fig.~\ref{fig3} shows the average SSR of different strategies versus SNR over 100 random channels with varied channel strength disparities and $ N_T=2 $. In both subfigures, consistent with Remark \ref{remark1}, the proposed CRS performs better than MU-LP and C-NOMA, which is due to the fact that the proposed CRS uses the common message not only to
serve as AN for impairing the decoding of the eavesdropper, but also to enhance the interference management among $ U_1 $ and $ U_2 $. When $\sigma_{h_1}^2=1,\sigma_{h_2}^2=0.3$, the average SSR gap between C-NOMA and CRS decreases, compared to $\sigma_{h_1}^2=1,\sigma_{h_2}^2=1$. It shows that C-NOMA is more suitable to the scenarios where the users experience a large disparity in channel strengths, though C-NOMA performs generally quite poorly compared to other schemes due to its inefficient use of multi-antenna degrees-of-freedom \cite{clerckx2019rate,mao2018rate}.
\section{Conclusion}
In this paper, we enhance PHY security in multi-antenna BC by using CRS. We aim at maximizing the SSR subject to transmit power constraint. The result, obtained by utilizing SCA-based algorithm to solve the non-convex problem, shows the proposed CRS secure transmission scheme achieves higher SSR than existing MU-LP and C-NOMA. The benefits of CRS comes from the use of the common message for dual purposes. It not only ensures the secure transmission of the private messages, but also the cooperation between the legitimate users. Therefore, CRS is a promising strategy to improve the PHY security in multi-antenna BC.
\section{Acknowledgement}
This work was supported by the National Key Research and Development Program under grant 2018YFB1801905. 

This work has been partially supported by the U.K. Engineering and Physical Sciences Research Council (EPSRC) under
grant EP/N015312/1, EP/R511547/1.

\ifCLASSOPTIONcaptionsoff
  \newpage
\fi

\small
\bibliographystyle{IEEEtran}
\bibliography{IEEEabrv,crs }
\bibliographystyle{unsrt}
\end{document}